\documentclass[preprint,journal]{vgtc}            

\onlineid{0}

\vgtccategory{Research}

\vgtcpapertype{please specify}

\title{\name{}: Guided Health Information Seeking from LLMs \\ via Knowledge Graph Integration}

\author{%
  Youfu Yan *, 
  \authororcid{Yu Hou *}{0009-0009-7184-6592}, 
  Yongkang Xiao, 
   Rui Zhang $^\ddag$, 
   \authororcid{Qianwen Wang $^\ddag$}{0000-0003-1728-4102}
}

\authorfooter{
  \item $\ast$ equal contribution, $\ddag$ co-corresponding authors
  \item Youfu Yan, and Qianwen Wang are with the Department of Computer Science and Engineering,  University of Minnesota, Twin Cities, MN, USA. E-mail: {yan00111, qianwen}@umn.edu
  \item Yu Hou, Yongkang Xiao, Rui Zhang are with the Medical School, University of Minnesota, Twin Cities, MN, USA.
  E-mail: {hou00127, xiao0290, zhan1386}@umn.edu  
}

\abstract{%
The increasing reliance on Large Language Models (LLMs) for health information seeking can pose severe risks due to the potential for misinformation and the complexity of these topics.
This paper introduces \name\, a visualization system that integrates LLMs with Knowledge Graphs (KG) to provide enhanced accuracy and structured exploration.
Specifically, for enhanced accuracy, \name\ extracts triples (\eg, entities and their relations) from LLM outputs and maps them into the validated information and supported evidence in external KGs.
For structured exploration, \name\ provides next-step recommendations based on the neighborhood of the currently explored entities in KGs, aiming to guide a comprehensive understanding without overlooking critical aspects.
To enable reasoning with both the structured data in KGs and the unstructured outputs from LLMs, \name\ conceptualizes the understanding of a subject as the gradual construction of graph visualization.
A progressive graph visualization is introduced to monitor past inquiries, and bridge the current query with the exploration history and next-step recommendations.
We demonstrate the effectiveness of our system via use cases and expert interviews.
}


\teaser{
  \centering
  \includegraphics[width=\linewidth, alt={A view of a city with buildings peeking out of the clouds.}]{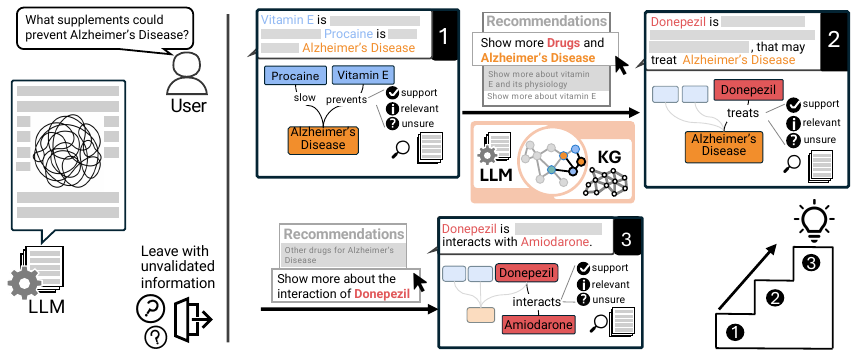}
  \caption{%
  	In contrast to traditional LLM question-answering (left), which often generate lengthy and unverified text, \name\ (right) leverages external knowledge graph (KG) to enhance health information seeking with LLM.
    \name\ provides validation through literature for accuracy, next-step recommendations for comprehensive exploration, and step-by-step graph visualization for a progressive understanding of the topic.
  }
  \label{fig:teaser}
}

\graphicspath{{figs/}{figures/}{pictures/}{images/}{./}} 

\usepackage{tabu}                      
\usepackage{booktabs}                  
\usepackage{lipsum}                    
\usepackage{mwe}                       

\usepackage{mathptmx}                  

\newcommand{\eg}{\textit{e.g.}}
\newcommand{\etal}{\textit{et al.}}

\usepackage{duckuments}
\usepackage{flushend}
\usepackage{pifont}

\usepackage{tikz}
\newcommand*\circled[1]{\tikz[baseline=(char.base)]{
            \node[shape=circle,fill,inner sep=0.5pt] (char) {\textcolor{white}{#1}};}}

\newcommand{\qianwen}[1]{#1}

\newcommand{\name}{\textsc{KnowNet}}

\begin{document}


\firstsection{Introduction}

\maketitle
Recently, Large Language Models (LLMs) have shown exceptional proficiency in a wide range of tasks and domains~\cite{fan2023bibliometric,de2023chatgpt,birhane2023science}, significantly transforming our approach to information seeking.
More and more people are now turning to LLMs to acquire the desired information on diverse topics.
Despite their advanced capabilities, the interactions with LLMs has been criticized for their insufficient factual accuracy~\cite{zhang2023siren, kaddour2023challenges}, lack of guidance in exploration~\cite{angert2023spellburst, yeh2022guide}, and inadequate support in representing intricate information structures~\cite{jiang2023graphologue}. 
These issues are particularly critical when seeking health-related information, where accuracy and clarity are paramount. 

One primary reason for these issues is that LLMs encode knowledge from the training corpus in the form of model parameters, which are difficult to interpret, validate, and align with users' cognitive processes.
As a result, one promising solution is to enhance LLMs with external knowledge that can be inspected and interpreted~\cite{lewis2020retrieval, pan2024unifying}.
Among various formats of external knowledge, knowledge graphs (KG) are attracting increasing attention due to their decisive knowledge representation and symbolic reasoning ability.
A knowledge graph stores structured knowledge as a network of entities and their relations, serving a broad range of domain applications including drug development~\cite{wang2022extending, huang2023zero}, children education~\cite{chen2023fairytalecqa, lee2022promptiverse}, and pedigree networks\cite{nobre2018lineage}.
Incorporating knowledge graphs in LLMs can provide structured knowledge when training the model~\cite{agarwal-etal-2021-knowledge}, retrieve pertinent information when responding to specific queries~\cite{lewis2020retrieval}, and offer evidence when reasoning the outputs~\cite{petroni2019language}. 
Such integration have been shown to significantly boost performance in domain-specific and knowledge-intensive tasks~\cite{pan2024unifying}.


While integrating knowledge graphs can markedly enhance the quality of LLM outputs, this enhancement is underutilized with current purely text-based interfaces.
 These interfaces, which typically rely on linear text formats like extended paragraphs, limit the user's ability to collect, organize, and synthesize information from the structured external knowledge, which has the potential to greatly facilitate the user's cognitive process.
Recently, a growing body of studies are exploring innovative interfaces for LLM via text-visualization coordination~\cite{zhang2023visar, jiang2023graphologue}, multi-level abstractions~\cite{suh2023sensecape}, node-link diagrams~\cite{angert2023spellburst}.
However, these studies mainly focused on enriching human-LLM interactions through prompt engineering, providing little discussion about the integration of external knowledge bases.
Meanwhile, many visual analytics methods have been proposed in the past decades to facilitate the interpretation of graphs for both domain-specific (\eg, biomedical knowledge graphs~\cite{wang2022extending, lex2013entourage}, neural network architectures~\cite{jin2022gnnlens, wang2019genealogy}) and general purposes~\cite{partl2016pathfinder, song2023gvqa, li2023networknarratives}.
But these visualization techniques mainly support information seeking through queries about graph structures (\eg, paths connecting nodes A and B, nodes with the highest degree) and cannot be directly applied to complex tasks that require iterative conversations.
There is a gap in effectively leveraging external knowledge graphs for more comprehensive and interactive visual interfaces for LLM.

This study proposes \name, a visualization system for health information retrieval by improving the traditional LLM-based with enhanced accuracy and structured exploration via integrating external KGs.
To enhance accuracy, we extracted triples (e.g., entities and their
relations) from LLM outputs and mapped them into the validated information and supported evidence in external KGs. 
For structured exploration, \name\ provides recommendations for further inquiry to help form a comprehensive understanding without overlooking critical aspects.
Considering a multi-step exploration might introduce information overwhelming, \name\ follows the focus+context design and proposes a progressive graph visualization to track previous inquiries, and connect this history with current queries and next-step recommendations. 
We demonstrate the effectiveness of our system via use cases and expert interviews
While we probed its capabilities initially within the context of dietary supplements (\eg, vitamins, minerals, herbs), an area where scientifically validated information is partially important given the prevalence of exaggerated claims and misinformation, the proposed approach is applicable to a wide array of applications.
The source code and documentations for \name\ are available at \url{https://visual-intelligence-umn.github.io/KNOWNET/}.

\section{Related Work}

\subsection{Improving the Usability of LLM}
As LLMs are increasingly employed, there is growing concern about their usability limitations stemming from both the inherent characteristics of LLMs and the design of current interfaces. Various research efforts are being undertaken to mitigate these issues.

One approach focuses on offering controllability through tailored guidance, especially via prompt designs.
Many studies have observed that, without further guidance, users tended to interact with LLMs opportunistically and struggle to make robust progress~\cite{zamfirescu2023johnny, feng2023promptmagician, brade2023promptify}. 
A list of tools has been proposed to guide the design of prompts for support LLM interactions.
For instance, AI Chains~\cite{wu2022aichain} provides an interactive system that chains LLM prompts and enables users to modify these chains in a modular way.
Promptify~\cite{brade2023promptify} and PromptMagician~\cite{feng2023promptmagician} utilizes a suggestion engine to help users quickly explore, craft, and organize diverse prompts.
Zamfirescu-Pereira~\etal~\cite{zamfirescu2023johnny} suggested strategies such as the use of example input/output pairs and the use of repetition within prompts.
Even though these studies have contributed valuable insights, it remains a challenge to design them effectively as there is no one-size-fits-all solution.
In a recent study, Su-Fang~\etal~\cite{yeh2022guide} investigated the effects of two guidance types and four guidance timings. Their study indicated there is no clear best choice for guidance type or timing, but depends on the specific goals of the guidance.

Meanwhile, a parallel research branch explores integrating graphical representations to complement traditional text interface. 
For instance, ChatGPT has included multiple plug-ins, such as Lucid GPT~\cite{lucid} and ChatGPT Diagrams~\cite{GPTDiagram}, that translate textual outputs into visual diagrams. 
However, these visual outputs are usually static images with limited or no interactivity.
To address this issue, Graphologue~\cite{jiang2023graphologue} constructs interactive graphical charts using novel prompt strategies.
Sensecape~\cite{suh2023sensecape} further enhanced these graphical charts by introducing hierarchical abstracts among which users can easily navigate.
The integration of graphical representation has also been demonstrated to improve task performance and user satisfaction in various domain applications, including argumentative writing~\cite{zhang2023visar} and creative coding~\cite{angert2023spellburst}.
Unlike prior studies that merely converted LLM text outputs to visual representations, our study aims to integrate external knowledge graphs to enhance the interactions with LLMs.

In spite of the great success of the above efforts, they cannot improve task performance in which LLMs have limited knowledge.
As a result, researchers proposed harnessing external knowledge to improve the output quality, known as retrieval-augmented generation (RAG)~\cite{lewis2020retrieval}.
RAG can enhance LLMs by querying relevant information from an external dataset for generating outputs, ensuring that the responses are grounded in retrieved evidence and include up-to-date knowledge.
Among different types of external knowledge bases, knowledge graphs are widely used due to their symbolic reasoning ability, as discussed in the survey paper of Pan \etal~\cite{pan2024unifying}.
For example, Ashby~\etal~\cite{Ashby2023Personalized} generated fluent and coherent dialogue for role-playing games by incorporating LLMs with a hand-crafted knowledge graph about the game world.
%
However, these studies mainly use external knowledge graphs for improving LLM outputs and do not explicitly elucidate how the integration can improve human-LLM interfaces, which is the main focus of our paper.



\subsection{Visualizing Graph-based Knowledge}
Graph visualization for knowledge communication has received extensive attention within the visualization community. A diverse array of visual analytics systems has been developed, covering diverse domains such as biology networks~\cite{wang2022extending, lex2013entourage}, neural networks~\cite{jin2022gnnlens, wang2019genealogy}, and pedigree networks\cite{nobre2018lineage}.
The principal challenge lies in effectively uncovering patterns within large graphs.
A variety of visualization techniques have been proposed for uncovering different types of patterns, including layout algorithms designed to expose communities within graphs~\cite{wang2020deepdrawing}, techniques for path querying and organization~\cite{partl2016pathfinder, wang2022extending}, and methods for conducting visual comparisons across different graphs~\cite{wang2019genealogy}.
For example, Wang \etal~\cite{wang2022extending} proposed a novel design, MetaMatrix, to help users organize and compare explanation paths in a biomedical knowledge graph at different levels of granularity.
In spite of the effectiveness of these methods, the challenges in graph exploration often extend beyond merely generating a static visualization but necessitate interaction techniques, leading to a set of popular interaction techniques such as focus + context and semantic zoom.
Focus + context techniques, such as fish-eye~\cite{sarkar1992graphicalfisheye}, display the object of interest in detail (focus) with an overview of surrounding information (context).
Semantic zoom~\cite{wiens2017semantic, lyi2021gosling} adjusts the visual representations dynamically based on the level of details.
Once patterns are identified, efficiently communicating them also poses another complex challenge, leading to investigations into the use of natural language interfaces and narrative techniques.
For instance, GVQA~\cite{song2023gvqa} facilitates the articulation of visual insights in graph visualizations through natural language. NetworkNarratives~\cite{li2023networknarratives} introduces semi-automatic data tours that elucidate network facts via slideshows with visualizations and textual annotations.

Our study is built upon previous studies on graph visualization, examining various visualization techniques for facilitating human-LLM interactions with the integration of external knowledge graphs.

\section{Informing the Design}

\name\ is designed to aid users in searching for health-related information by seamlessly integrating LLM outputs with external KG. 
It targets individuals who require access to such information and possess the ability to grasp complex medical concepts and interpret research findings. 
Primary users of this tool include medical researcher, health science students, and patient advocates.

\subsection{Design Requirements}
We identify the design challenge based on discussion with domain experts and review of literature, following the practices in \cite{wang2020discrilens, cheng2022polyphony, wang2021threadstates}. 
\qianwen{Specifically, three authors are experts in computational health sciences with extensive experience, including a professor with 15 years of research in health AI, knowledge graphs, and clinical NLP; a postdoc researcher with 6 years of experience in knowledge graphs and EHR data analysis; and a PhD candidate with 3 years of experience in knowledge graphs and health informatics.
}

\qianwen{
Four authors, including two domain and two visualization experts, first collaboratively built a list of relevant papers that examined, applied, or improved LLMs for various information-seeking tasks, including but not limited to writing~\cite{zhang2023visar, lee2022coauthor}, coding~\cite{zamfirescu2023johnny, angert2023spellburst}, and healthcare~\cite{yang2023harnessing}.
Each paper was then independently reviewed by at least two authors. Drawing on the findings from the literature review, the author team held weekly meetings to refine the design requirements, update the design, and test the developed tool. 
Through this iterative process, we identified five key design challenges.
}

\begin{enumerate}[leftmargin=*, label=\textbf{C.\arabic*}]
\item \label{challenge:linear} \textbf{Linear Response to Hierarchical Information:} LLMs, by design, generate responses in a linear fashion, presenting information in a single continuous stream. 
However, the structure of knowledge itself is intrinsically marked by complex and multifaceted relationships among concepts.
Taking a common dietary supplementary vitamin D as an example. 
At a basic level, understanding vitamin D involves understanding its various forms (\eg, D2, D3), sources (\eg, sunlight, food), and its role in the human body, such as supporting bone health and immune function. 
Delving deeper reveals a web of interconnected details that spans nutrition, biochemistry, and public health, such as its involvement in calcium absorption and the implications of its deficiency.
As a result, the current linear presentation off LLMs can represent users to grasp the complex structure and navigate intricate topics~\cite{suh2023sensecape,jiang2023graphologue,zhang2023visar}.

\item \label{challenge:limited_verification} \textbf{Limited Support for Verification:} 
LLMs often suffer from hallucination, wherein an LLM generates incorrect information but presents it as it was a fact, which can potentially deceive users who lack the expertise to assess the information accuracy.
This can lead to potential deception among users who may not have prior knowledge to evaluate the accuracy of the information.
Several LLM-based chatbots have integrated internet search to alleviate this issue, but the referred online resources have various reliability.
In a critical context, such as medicine and healthcare, users have expressed a preference for AI tools that function similarly to knowledgeable colleagues that can reference reliable evidence, such as biomedical research, to support their responses \cite{yang2023harnessing}.

\item  \label{challenge:overwhelming} \textbf{Information Overload:} 
LLMs are often designed to generate a verbose, long-form answers that includes extensive information~\cite{lewis2020retrieval, lewis2019bart}.
Users can be overloaded, feel that \textit{``there is too much to read''}, and find it challenging to efficiently interpret the response and comprehend the underlying reasoning chain~\cite{lee2023dapie}.
Additionally, these long-term responses tend to include redundant information and utilize pompous language \cite{chen2023chatgpt}, which further hampers the user's comprehension process.

\item \label{challenge:lack_guidance} \textbf{Lack of Exploration Guidance:} LLMs excel at supporting free-form exploration, with their extensive knowledge base enabling them to answer a wide variety of user questions. 
However, their propensity for broad exploration can sometimes be overwhelming. 
Without guidance, users often find themselves confused about \textit{"what to ask next, and how?"} due to the sheer volume of content available for inquiry~\cite{yeh2022guide, zamfirescu2023johnny}.
This issue is especially salient for complex topics that cannot be fully addressed with one single question, but requires iterative follow-up conversations \cite{kim2023cells}.

\item \label{challenge:goal_absence} \textbf{Absence of Goal-Achieving Indicator:} 
LLMs typically do not have an inherent mechanism for aligning its outputs with the user's goal of understanding a specific topic, which often varies from person to person.
Users might find themselves navigating through an abundance of information without a clear sense of \textit{``how much more information is needed''} for forming a desired level of understanding. 
Often, they might quit the information seeking after asking the one or two questions (\eg, whether a drug can be used for a disease), overlooking other important related information (\eg, the side effort of this drug).

\end{enumerate}

\begin{figure}
    \centering
    \includegraphics[width=0.9\linewidth]{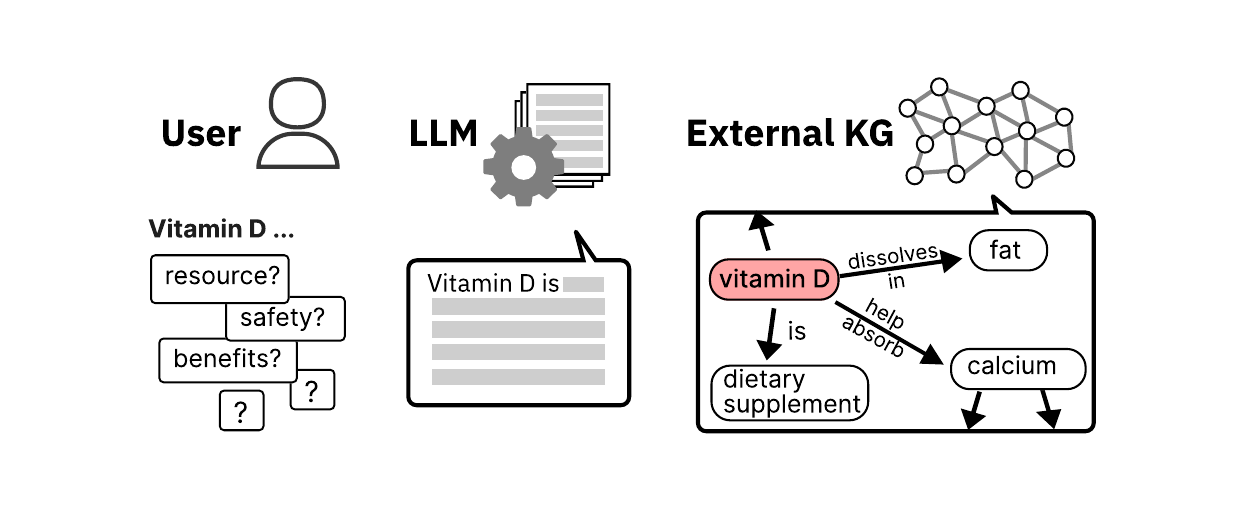}
    \caption{\name\ is designed to support the communication among three distinct forms of knowledge: the knowledge users apply in their reasoning process, the knowledge contained within LLMs, and the knowledge stored in KGs.}
    \label{fig:three-spaces}
\end{figure}

\subsection{Three Forms of Knowledge}

\name\ aims to provide a systematic solution to the identified challenges in information seeking via the integration of KGs.
Therefore, \name\ needs to enable the communication among three distinct forms of knowledge, the one used in users' reasoning process, the one stored in LLM, and the one contained in KG (\cref{fig:three-spaces}). 
We consider the process of understanding an object as progressively forming a knowledge graph about it. 
The external knowledge graph serves as a scaffold for organizing information, assessing LLM response, guiding exploration, and tracking the exploration progress.

This idea is driven by the observation that knowledge graphs, by nature, align with the principles of symbolic cognitive modeling, a classic approach for modeling human cognitive process~\cite{wilson2001encyclopedia}.
Knowledge graphs represent information in a structured, symbolized format, using nodes to represent entities (\eg, objects) and edges to represent the relationships between these entities. 
This structured representation mirrors the symbolic cognitive model, which posits that human cognition operates through the manipulation of discrete symbols, represented by the nodes and edges in knowledge graphs.
For example, understanding the node \texttt{[Paris]} involves recognizing its connection through a \texttt{[capital\_of]} edge to the node \texttt{[France]}. 
By actively, progressively constructing a knowledge graph around the object of interest, users engage in a symbolic manipulation process, drawing connections and making inferences that enhance their understanding in a manner akin to typical human reasoning.

\begin{figure}
    \centering
    \includegraphics[width=\linewidth]{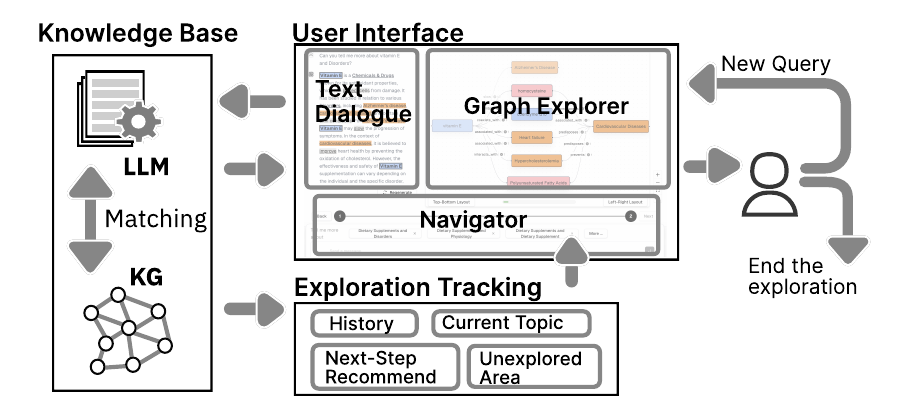}
    \caption{\textbf{System Overview.}}
    \label{fig:overview}
\end{figure}

\subsection{System Overview}
\name\ consists of three main modules, a knowledge base, an exploration tracking module, and a user interface, as shown in \cref{fig:overview}. 
The knowledge base merges LLM and KG capabilities to deliver structured and verified responses to user inquiries.
Based on the current query, \name\ extracts pertinent information from the KG and generates recommendations for next-step exploration, which are then sent to the exploration tracking module. 
The user interface enables users to interact with both the knowledge base and the exploration tracking module.
Users can validate responses, explore recommendations, and adjust their exploration goals, determining whether to proceed with the current conversation.

\section{Integrating Knowledge Graph}
\begin{figure}
    \centering
    \includegraphics[width=\linewidth]{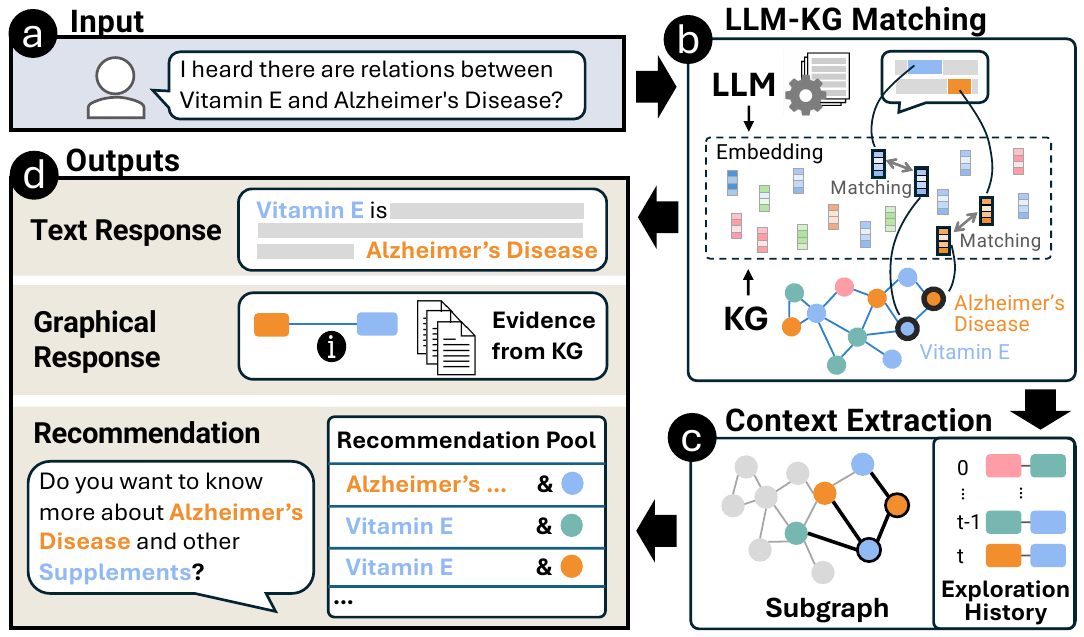}
    \caption{
    \qianwen{\textbf{An overview of the back-end}. \name\ accepts text inputs (a), maps entities in the LLM response to corresponding nodes in the KG based on their embeddings (b), and identifies related entities in the KG neighborhood to generate recommendations based on user exploration histories (c). Finally, \name\ outputs text responses and visualizations that organize the main relationships, provide evidence from the KG, and suggest next-step recommendations (d).}
    }
    \label{fig:KG}
\end{figure}

In this section,  we outline our approach for integrating external KG with LLMs to address the identified design challenges (\ref{challenge:linear}-\ref{challenge:goal_absence}). \qianwen{\cref{fig:KG} shows the overview of the back-end structure.}
We focus on using KG to enhance user interactions with LLMs, complementing existing studies that have used KGs to improve the pre-training, fine-tuning, and inference processes of LLMs.


For the LLM, we selected GPT4 as it is the state-of-the-art method.
For the KG, we utilized ADInt\cite{xiao2023repurposing}, a KG we developed in earlier studies by extracting information on Non-pharmaceutical interventions and  Alzheimer's Disease from biomedical literature. 
This comprehensive KG comprises 162,212 nodes (15 types of entities such as drug, disease, symptoms) and 1,017,284 edges (the relation between entities), incorporating a total of 754,224 pieces of scientific literature.
ADInt serves as a testbed in this study and the proposed method is designed for easy adaptation with other knowledge graphs, especially considering the broad availability of knowledge graph datasets across various domains~\cite{pan2024unifying}.

\subsection{Respond to current query (\ref{challenge:linear}, \ref{challenge:limited_verification})}
\label{subsec:response}

After a user poses a question, \name\ responds by combining the LLM outputs and the related information from KG \qianwen{(\cref{fig:KG}(b))}. 

\qianwen{The LLM will first determine whether the user question falls within the scope of the KG. If the query is outside the KG's scope, \name will function the same way as a standard LLM chat. Otherwise, the LLM will be prompted to annotate triples for extracting structured information from the unstructured text (\ref{challenge:linear}). 
}
One triple consist of two entities and their relation. 
Inspired by the prompting strategies in Graphologue\cite{jiang2023graphologue}, we prompted the LLM to assign a unique identifier for each entity ($n_1, n_2, ...$) and their relations ($r_1, r_2, ...$) simultaneously during text generation. 
For example, in the response
\textit{``[fish oil](\$n1) is known for [containing](\$r1, \$n1, \$n2) a rich content of [Omega-3 fatty acids](\$n2)''}, GPT identifies one triple with two entities: \textit{fish oil}, denoted as $n_1$, and \textit{Omega-3 fatty acids}, denoted as $n_2$, and their relation, \textit{containing}, denoted as $r_1$.

We then match these triples with the KG to extract related literature and provide assistance in verification (\ref{challenge:limited_verification}).
\qianwen{Specifically, we generate embedding vectors for the nodes in the KG and the entities in the LLM outputs using the same embedding model, the \textit{text-embedding-ada-002} model from OpenAI.}
Given a new LLM triple $(n_i, r, n_k)$, we identify the corresponding KG nodes by calculating and comparing the cosine similarity of their embeddings to those of the nodes within the triple, \qianwen{as shown in \cref{fig:KG}(b)}.
We then search in the KG to find whether the two entities are connected via either one- or two-hop paths.
\qianwen{These paths connecting entities are derived from mining biomedical literature, providing useful insights for interpreting and validating LLM responses. 
New knowledge can be integrated into \name by updating the KG through the insertion of new entities and paths.
}
As noted in previous studies \cite{yang2023harnessing, wang2022extending}, the biomedical literature is a preferred resource to calibrate user trust with AI tools.
It is important to note that KG paths between two entities do not always imply the same relation in the LLM triple. 
This discrepancy is particularly pronounced due to the ambiguity of text. 
For example, ``slow the progress'' does not equal to ``treating'' a disease.
Similarly to node matching, we applied text embedding to compare relations suggested by LLM with those identified in the KG. 
\qianwen{We consider these relations to be equivalent if the cosine similarity of their embedding vectors exceeds a certain threshold, which may need refinement based on the specific KGs and embedding methods used. In our implementation, we set the threshold at 0.94 based on experiments.}


\begin{figure*}
    \centering
    \includegraphics[width=\linewidth]{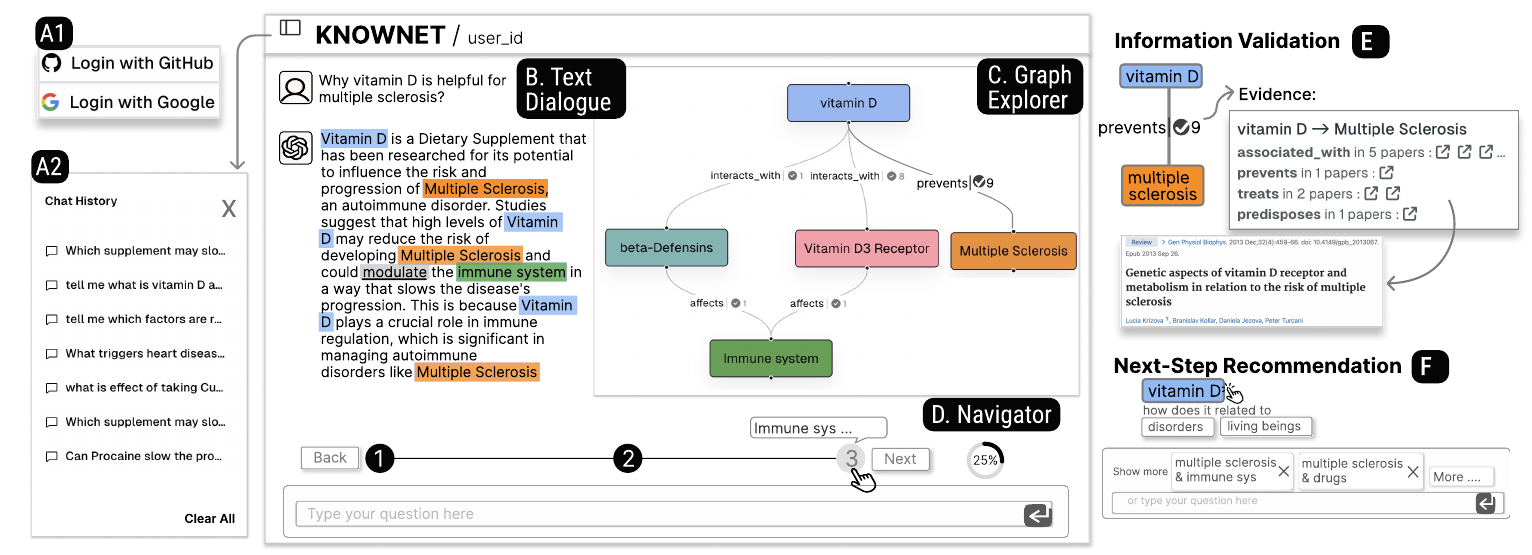}
    \caption{\textbf{Interface of \name}. Users can log into the tool via GitHub or Google account and securely store their chat history (A1-A2). The main interface consists of a \textit{Text Dialogue} (B), a \textit{Graphical Explorer} (C), and a \textit{Navigator} (D). Information validation (E) and next-step recommendations (F) are provided to facilitate the exploration. }
    \label{fig:interface}
\end{figure*}

\subsection{Recommend further explorations (\ref{challenge:lack_guidance}, \ref{challenge:goal_absence})}

\name\ generates recommendations considering both the KG structured neighborhood and the user exploration history (\ref{challenge:lack_guidance}), \qianwen{as shown in \cref{fig:KG}(c)}.
Even though we can generate next-step recommendations by prompting GPT, \eg, \textit{``please suggest relevant questions for further exploration.''}, it offers limited control to the users.
Therefore, we used the structured information in KG to provide customizable recommendation.

We categorize user queries $Q$ into two main categories based on the format: the relation between a node and a node type , or the relation between a node and another node. 
$$Q=(N, \{T|N'\})$$ 
where $T$ represent node types and $N$ represent nodes.
For example, with vitamin E as the node, query examples include ``which disorders can vitamin E improve'' (a node and a node type), ``is vitamin E helpful for Alzheimer's disease'' (a node and another node).
Our recommendations are crafted following such query patterns.


We model user state based on their exploration history following the n-context analysis outlined by Milo and Somech \cite{milo2018next}.
We denote the user's current state using the previous queries posed by users, 
$$context_t=(q_0, q_1, ..., q_t)$$ 
where $q_0, q_1, ..., q_t \in Q $ represent the query at corresponding time steps $0,1, ...t$.
With this modeling, we are able to map users' current state as a sequence of nodes or node types and map them into the KG, which serves as a foundation for recommending queries for further exploration.

To model of the goal of an information seeking process (\ref{challenge:goal_absence}), we extract the one-hop neighbors of the entities mentioned in the user's initial query from the KG, assuming the initial query reflects the primary objective of their information search. 
In other words, 
$$Subgraph_Q=\bigcup_{n_i \in {n_1, n_2, .., n_k}} Neighbor(n_i)$$ 
where ${n_1, n_2, n_k} \in N$ are the entities mentioned by the users in the initial query, $Subgraph_Q$ is the neighborhood in KG that contains potential subjects for exploration.
$$ Recommendations = \bigcup_{n_i \in context \land \{n_j|t_k\} \in Subgraph_Q } (n_i, n_j|t_k) $$
We then use a rule-based template to convert the candidate queries into natural language questions.


This recommendation mechanism may not always align with unique goal of individual users in their information seeking.
For example, users may only care about the benefits of vitamin E but not its physiology aspect, even though both aspects are treated as equally important in the KG.
To address this issue, we update the recommendation pool based on two types of user feedback. First, users can remove recommendations that don't capture their interest. Second, they can pose new, highly relevant questions that are not in the recommendations. 
The recommendation pool is dynamically updated, removing these less relevant queries and incorporating new suggestions by updating the $Sgubgraph_Q$ with new entities in user-added questions.

\section{Visual Interface}

As shown in \cref{fig:interface}, the interface of \name\ consists of three main components, a \textit{Text Dialogue} (B), a \textit{Graphical Explorer} (C), and a \textit{Navigator} (D).
The \textit{Text Dialogue} presents the text response to the current query, highlighting related entities and their relations for seamless integration with the \textit{Graph Explorer}.
The \textit{Graph Explorer} not only represents the text responses as graphical representations but also enables easy validation of these responses by matching the information with evidence in the KG.
Previous exploration is also summarized in the \textit{Graph Explorer}, enabling users to easily connect the current query to previous explorations and next-step recommendations.
The \textit{Graph Explorer} is updated according to the \textit{Navigator} module.
The \textit{Navigator} allows users to revisit previous queries, gauge their progress in exploration, and submit queries based on recommendations.
Although we describe the interface as three separate modules for clarity, they are seamlessly integrated with one another.

\subsection{Information Seeking with \name}
We introduce the components and interactions in \name\, demonstrating how it can guide users in information seeking and solve the design challenges.

\subsubsection{Connecting LLM with KG}

To start with, users will post a question about an object of interest, such as \textit{``what are the benefits of taking vitamin D?''}.
As outlined in \cref{subsec:response}, we extract triples (\eg, entities and their relationships) from the GPT outputs and match them with KG data. 
During the streaming generation process, identified entities are highlighted in gray and their relations are underlined (\cref{fig:interface}(B)). 

\qianwen{
Upon successfully matching these entities within the KG, the identified entities and their relations are updated to \textit{Graph Explorer} as a node-link diagram.
We apply color coding to indicate differentiate node types, such as drugs, diseases, and physiological aspects. 
Entities in the Text Dialogue and the Graph Explorer share the same color coding and are synchronized, as shown in \cref{fig:interface}(C). 
Hovering over an entity in one view will highlight the corresponding entity in the other view.
This coordination between \textit{Text Dialogue} and \textit{Graph Explorer} enables an intuitive presentation of nonlinear entity relations (\ref{challenge:linear}) and provides a concise representation of the main message in potentially long texts (\ref{challenge:overwhelming}).
}
Unlike Graphologue~\cite{jiang2023graphologue}, we choose not to update the graph simultaneously for two main reasons. 
First, the sheer size of the KG may introduce delays. 
Second, the complex structure of responses to medical queries makes simultaneous monitoring of both text and graph potentially overwhelming for users.

Users can click on the node in \textit{Graph Explorer} to highlight the corresponding entities in the GPT outputs.
Users can also move nodes through drag-and-drop to better organize the node layout.

\begin{figure}
    \centering
    \includegraphics[width=\linewidth]{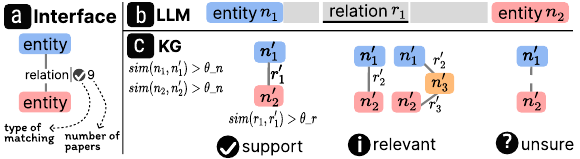}
    \caption{\textbf{Edge labels}. \name\ suggests three different edge labels, support, relevant, and unsure, based on the matching between KG and LLM. $sim(n_1, n_1')$ indicates the cosine similarity between the the entity identified in LLM ($n_1$) and the nodes in KG ($n_1'$). $\theta_n$ and $\theta_r$ are the thresholds for entity and relation matching. }
    \label{fig:edge-label}
\end{figure}

\subsubsection{Verification with Evidence}

\qianwen{We provide various edge labels and pop-up details in \textit{Graph Explorer} to aid users in evaluating the quality of a relation (\ref{challenge:limited_verification})}. 
As shown in \cref{fig:edge-label}(a), each label consists of three components: the name of the relation, one of three classifications (Support, Relevant, or Unsure), and the number of related literature found in the KG. 
Clicking on the label opens a pop-up window that lists the relevant literature and allows users to dive deeper into these evidences, as shown in \cref{fig:interface}(E).

We classify the quality of the relations as support, relevant, and unsure, based on the matching between KG and LLM (\cref{fig:edge-label}(b-c)).
The support label \circled{\ding{51}} is assigned when GPT-mentioned relations can be directly corroborated with evidence found in the KG.
The relevant label \circled{i} will be assigned in two scenarios. 
First, this label will be assigned if a similar but not identical relation is found in the KG compared to what GPT mentioned. 
For example, GPT might suggest that a drug can slow the progression of a condition, but the KG shows the drug can prevent a condition. 
Secondly, this label will be assigned if no direct link exists between two entities, but a two-hop path is discoverable in the KG, suggesting potential underlying mechanisms that could support the relation.
The unsure label \circled{?} will be assigned to relations to which neither the Support nor Relevant labels apply.
In addition to adding the question icons, we distinguish unsure relations with a dashed line, enhancing user understanding of the relation's credibility.

\subsubsection{Next Step Recommendation}

\qianwen{\name\ offers next-step recommendations in both the \textit{Navigator} and the \textit{Graph Explorer} to encourage further exploration (\ref{challenge:lack_guidance}), as shown in \cref{fig:interface}(F).}
In the \textit{Navigator}, recommendations are displayed above the input box, with the top three suggestions presented as individual buttons. 
Additional recommendations become visible upon hovering over the ``More'' button. Meanwhile, in the \textit{Graph Explorer}, recommendations are linked to relevant nodes in the graph, aiding users in integrating these suggestions into the context of their ongoing exploration. 
Clicking on a button will automatically submit the corresponding recommended query to \name.

There may be instances where the recommendations do not fully align with the unique goals of individual users, particularly when there is a limited chat history available.
In such cases, users have the option to dismiss recommendations by clicking on the cross icon or to enter their own query into the input box.

\begin{figure}
    \centering
    \includegraphics[width=\linewidth]{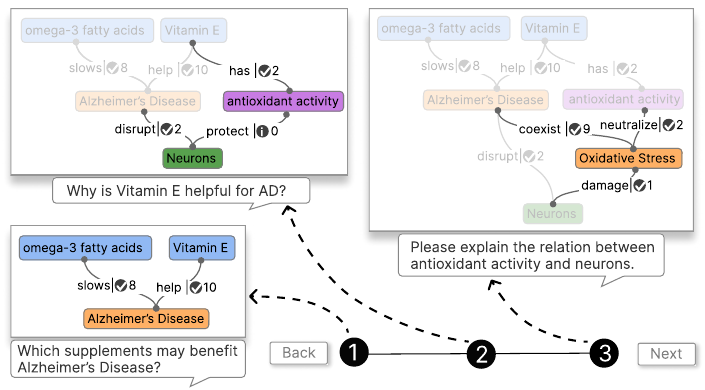}
    \vspace{-6mm}
    \caption{\textbf{Step by step exploration}. In \name, users seek information while progressively constructing a graph about it. With each step, newly added edges and nodes are highlighted, whereas elements from previous steps are faded. This design aims to help users concentrate on the current query while still retaining an awareness of the overall context.}
    \label{fig:steps}
\end{figure}

\subsubsection{Progress Tracking}

\qianwen{As users ask questions and progressively build a comprehensive understanding, the process can also result in an overwhelming amount of information, causing them to lose track of the query process~\cite{wang2018narvis}
Therefore, 
we provide a \textit{Navigator} that both structurally organizes exploration histories (\ref{challenge:overwhelming}) and indicates the ratio of explored area within the knowledge graph neighborhood (\ref{challenge:goal_absence}).
}


In the \textit{Navigator} (\cref{fig:interface}(D)),
each dot in the stepper indicates a query.
\qianwen{\name\ uses a linear Navigator to prioritize a familiar and straightforward navigation experience for users.}
Hovering on a dot will reveal the corresponding query text in a tooltip, and clicking on the dot navigates to the page of this query.
In the query's page, \textit{Text Dialogue} shows the text response to this query and \textit{Graph Explorer} will highlight the relevant nodes and their relations, fading previous explorations, and hiding explorations after this query, as shown in \cref{fig:steps}.
Compared to the scrolling-down layout of traditional conversational chatbot, this design enable users to focus one specific query without losing the context \qianwen{(\ref{challenge:overwhelming})}, embodying the focus+context design principle prevalent in visualization.

Additionally, a circular progress bar on the stepper's right side shows the proportion of the neighboring area in KG explored by the user \qianwen{(\ref{challenge:goal_absence})}. 
This target neighborhood is dynamically updated as users dismiss recommendations or introduce new queries, ensuring a tailored and manageable exploration experience.

\subsection{Account Management and Chat Histories}

\name\ allows users to sign in using either their GitHub or Google accounts and requires an OpenAI API key at the sign-in page. This integration is achieved through NextAuth.js\cite{_2024_nextauthjs}, a library for Next.js that provides a simple and secure solution for handling authentication in server-side rendering and static site generation applications. By using OAuth providers like GitHub and Google, we offer users a convenient way to access \name\ and maintain user data security. 

Chat histories are stored on Vercel KV \cite{_2024_VercelKV}, a durable Redis database that enables the storage and retrieval of JSON data. Storing chat histories ensure that users' conversation histories are preserved across sessions. 
Users can revisit previous conversations and continue their research from where they left off. 
More importantly, these histories can be used for generating a more personalized recommendation pool.


Overall, the account management and chat histories in \name\ enhance the usability and effectiveness of the system, providing users with a more personalized and continuous research experience.

\subsection{Implementation}
The implementation of \name\ involves a front-end for user interaction and a back-end for data processing and AI model integration. The front-end is developed using Next.js\cite{nextjs_2024_nextjs} to enable server-side rendering and static site generation. The chat functionality is powered by the Vercel AI SDK\cite{nextjs_2024_nextjs}, which provides a streamlined way to integrate AI chat models, such as OpenAI's GPT-4, into the application. Chat history and sessions are stored on Vercel KV\cite{nextjs_2024_nextjs}, ensuring that the user's conversation history is preserved across sessions. For authentication, NextAuth.js\cite{_2024_nextauthjs} is integrated into the application, providing a simple and secure solution for managing user authentication and session management.
The back-end of is built using Flask \cite{grinberg2018flask}, a lightweight Python web framework. Flask serves as the bridge between the front-end and the graph database, where Neo4j\cite{neo4j_2024_neo4j} is utilized to store and retrieve knowledge graph data. The source code and documentation for \name\ are available at \url{https://visual-intelligence-umn.github.io/KNOWNET/}.

\begin{figure*}[!ht]
    \centering
    \includegraphics[width=\linewidth]{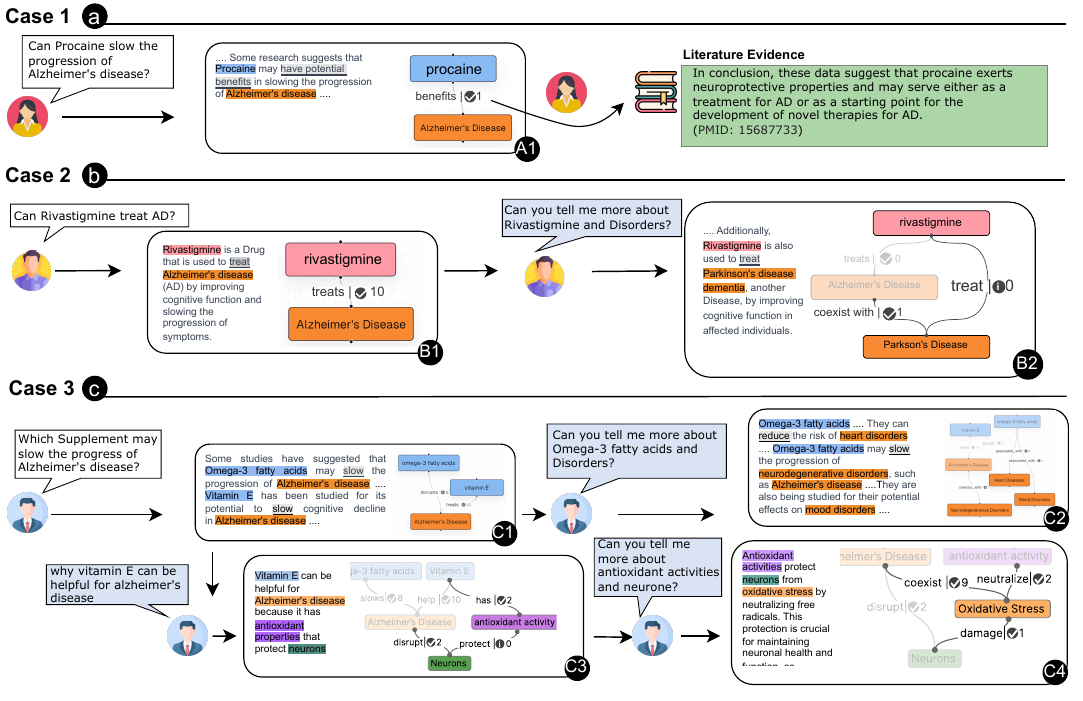}
    \caption{\textbf{Three typical use cases}. The blue chat bubble represents recommended questions, and the white chat bubble represents free questions. }
    \label{fig:usercase}
\end{figure*}

\section{Evaluation}
A total of 15 use cases were conducted to showcase the utility and usability of \name, performed by the three domain experts on our author team. 
Each use case involved a series of dialogues centered on one specific question related to dietary supplements.

In this section, we first present three representative use cases from the set of 15. 
Following this, we discuss our observations and the errors encountered across all use cases. 
Lastly, we conducted expert interviews with two additional domain experts who are not paper authors, further enriching our analysis and findings.


\subsection{Use Cases}
\label{subsec:usecase}

We first report three representative cases that simulated different usage scenarios, as shown in \cref{fig:usercase}, and then discuss our observations from conducting the 15 cases.


\subsubsection{Case One: Verification via Literature}
This case (\cref{fig:usercase}(a)) demonstrates how \name\ can help users verify information. The user inquired about the impact of one specific dietary supplement on Alzheimer's disease. 
In response to the query \textit{``Can Procaine slow the progression of Alzheimer's disease?''}, \name\ stated that: \textit{``Procaine may have potential benefits in slowing the progression of Alzheimer's disease.} 
The retrieved triple ([Procaine]-[prevents]->[Alzheimer's Disease]) was incorporated into the \textit{Graph Explorer}, \qianwen{with a support label \circled{\ding{51}} and corresponding literature evidence.}
The user can then read the corresponding paper to seek further information about the relation between Procaine and Alzheimer's disease.


\subsubsection{Case Two: Integrating Information from LLM and KG}
This case (\cref{fig:usercase}(b)) shows how \name\ integrates information from both LLM and KG for effective information seeking. The user investigated different drugs by beginning with, ``Can rivastigmine treat AD?'' stemming from a recent drug advertisement they encountered.
\name\ confirmed the treatment possibility \qianwen{with a support label \circled{\ding{51}}} and provided relevant evidence to support this claim.
Curious for more details about this supplement, the user pursued a recommended query, \textit{``Can you tell me more about Rivastigmine and Disorders?''}. 
\name\ indicated that \textit{``Rivastigmine is also used to treat Parkinson's disease dementia.''}
\qianwen{However, \name\ did not find a direct edge between Parkinson's disease and Rivastigmine, resulting in number 0 on the edge label. 
On the other hand, the KG found that Parkinson's disease can be connected to Rivastigmine via the node Alzheimer's disease (B2), leading to a relevant label \circled{\textbf{i}} rather than an unsure label \circled{?}. 
Users further examined the two-hop path via Alzheimer's disease and believed it cannot help verify the relation between Parkinson's disease and Rivastigmine. 
This observation suggested potential inaccuracies in the information provided by the LLM.
}

\subsubsection{Case Three: Guided Exploration}
This case (\cref{fig:usercase}(c)) shows how \name\ supports guided exploration with recommendations and progressive visualizations. The user started with asking \textit{``Which supplement may slow the progression of Alzheimer's disease?''} 
\name\ suggested Omega-3 fatty acids and vitamin E, noting that \textit{``Omega-3 fatty acids may slow the progression of Alzheimer's disease''} and \textit{``Vitamin E has been studied for its potential to slow cognitive decline in Alzheimer's disease''}. 
The retrieved triples ([Omega-3 fatty acids]-[affects]->[Alzheimer's Disease]) and ([Vitamin E]-[affects]->[Alzheimer's Disease]) were also updated in the graph \qianwen{with support edge labels \circled{\ding{51}}}. 
Based on the previous answer, the user was interested in one supplement, Omega-3 fatty acids, proposed by \name, and want to know which other disorders can benefit from taking this supplement. This question was exactly recommended by \name\ and displayed above the input box as ``Omega-3 fatty acids and Disorders.''
The user clicked on the responding button and \name\ generated responses accordingly: \textit{omega-3 fatty acids can be used for reducing heart disorders, slowing down the progression of neurodegenerative disorders and managing mood disorders}. 
\qianwen{
These relations were also updated to the \textit{Graph Explorer} with supporting edge labels \circled{\ding{51}}. 
}
In \textit{Graph Explorer}, nodes relevant to the current query (i.e., mood disorders, neurodegenerative disorders) are highlighted, aiding users in maintaining focus without getting overwhelmed by other nodes in the graph.

The user then revisited earlier steps to inquire about another recommended supplement, vitamin E.
The user selected the recommended question from \name, \textit{``why vitamin E can be helpful for Alzheimer's?''}.
\name\ responded with ``Antioxidant properties'' and their effect on ``neurons'' to explain why vitamin E was recommended.
As show in \cref{fig:usercase}(C3),
\qianwen{
a relevant label \circled{i} on the edge indicated that there are no direct edges, but there are two-hop paths in the KG connecting the two nodes.
}
Consequently, the user pursued further clarification by selecting a \name\ suggested question, asking, \textit{``Can you tell me more about Antioxidant properties and neurons?''}
The answer introduced a new concept (\cref{fig:usercase}(C4)), ``Oxidative stress'', \qianwen{which was linked to both ``Antioxidant properties'' and ``neurons'' with support edge labels \circled{\ding{51}}. These links helped validate the relation between ``Antioxidant properties'' and ``neurons''.
}

\
\subsubsection{Observations}

\noindent
\textbf{Complementarity between LLM and KG:}
LLMs can add context and details to the abstract and inflexible structures of KG, while KGs can enhance LLMs with accuracy and structured knowledge.
The use cases show how LLM and KG can mutually enhance each other's capabilities. 

First, LLM and KG can complement each other by providing information from different resources.
An instance of KG enhancing LLM occurs in the first Q\&A of Case 1, where the LLM's response regarding the efficacy of Procaine in Alzheimer's treatment was marked by uncertainty, indicated by phrases like \textit{``some research suggests that Procaine may have potential benefits in...''} 
Here, KG plays a crucial role in affirming this connection, supplementing the claim with supporting literature.
Conversely, an example of LLM augmenting KG is observed in the second Q\&A of Case 2.  
The LLM introduces a triple ([Rivastigmine]-[treats]-> [Parkinson's disease dementia]) that is not previously recorded in the KG. 
Therefore, the KG leveraged a two-hop path ([Rivastigmine]-[treats]->[Alzheimer's disease]-[coexists with]->[Parkinson's disease dementia]) to suggest Rivastigmine's applicability in treating a closely related disease.
This strategy is frequently employed in drug repurposing~\cite{wang2022extending}.
Although the validity of this relation requires further verification, it underscores the significant potential of LLMs to enhance the quality of the KG.

Second, the integration of text and graphical representation through LLM and KG collaboration enables varied perspectives for data interpretation. The textual descriptions generated by LLMs can offer nuanced explanations and background information.
Take the Omega-3 fatty acids in Case 3 as an example, the text description included information about its resource \textit{``rich in fish oil''} and main properties \textit{``anti-inflammatory and neuroprotective properties''}.
At the same time, the visual graphs provide a clear, immediate summary of the connection between Omega-3 fatty acids and Alzheimer's disease.
More importantly, the abstract nature of the graph facilitates easier recall of previous explorations, offering a contextual backdrop for understanding the current query. In contrast, extensive text segments tend to focus narrowly on information pertinent to the immediate query.

\begin{figure}
    \centering
    \includegraphics[width=\linewidth]{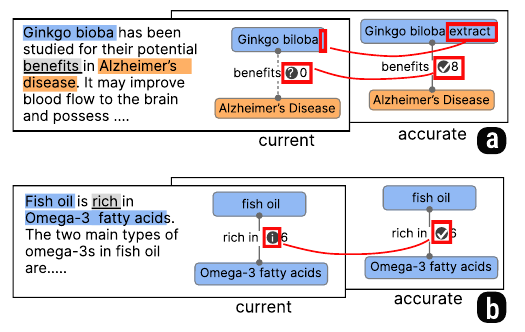}
    \caption{\qianwen{
    \textbf{Error cases}: (a) The LLM uses vague terms that can't be accurately mapped to domain-specific terms in the KG; (b) The KG has a limited scope and does not include common-sense knowledge that is not explicitly detailed in research articles.}
    }
    \label{fig:error-cases}
\end{figure}


\noindent
\textbf{Common Issues:}
\qianwen{
Analysis of the 15 cases also revealed various issues in the integration of LLM and KG, which we summarize as below.
}

First, aligning general language terms from LLM output to these standardized terms in KG can lead to inaccuracies. 
The inherent ambiguity of natural language can lead to LLM-generated entity and relation names being broad and nonspecific. 
At the same time, domains requiring high precision often rely on standardized terminology for clarity and consistency, such as the Unified Medical Language System (UMLS)\cite{bodenreider2004unified} used in \name. 
It is possible that a term used by an LLM is vague and can match multiple nodes in KG.
\qianwen{
For example, as shown in \cref{fig:error-cases}(a), even though the KG contains evidence to support the relation [Ginkgo biloba extract]->[benefit]->[Alzheimer’s Disease], LLM use a broder term ``Ginkgo biloba'', which was matched to the node ``Ginkgo biloba'' rather than ``Ginkgo biloba extract'' in KG. This results in a unsure label \circled{?} for the relation.
%
}


\qianwen{
Second, KGs are usually limited to specific domains and cannot be used to validate information outside those domains. 
In the current implementation, the \textit{Graph Explorer} will not be updated if LLM determines that a question falls outside the KG's scope.
However, this approach can fall short for common-sense knowledge within a field, which is often not explicitly detailed in research articles.
For example, the widely recognized fact that \textit{``fish oil is rich in Omega-3 fatty acids''} is labeled as relevant \circled{i} rather than support \circled{\ding{51}}, as shown in \cref{fig:error-cases}(b).
}

\qianwen{
Third, GPT4 tend to be overly cautious, often using vague terms and occasionally refusing to answer certain questions. For instance, when queried about the connection between Procaine and Alzheimer's Disease, an LLM responded with the statement, ``Procaine may have potential benefits in slowing the progression of Alzheimer's Disease.'' 
This cautious approach of GPT can sometimes make \name\ unresponsive to user questions. However, it can also result in fewer instances than expected of uncertain edge labels \circled{?} in the use cases.
}

Fourth, even though integrating the additional graphical format can significantly address the issues in traditional linear textual description, not all knowledge can be effectively represented as graphs.
For example, consider the sentence \textit{``Oxidative stress is related to the process of neuronal damage, as it involves the accumulation of harmful reactive oxygen species that can damage neurons''}.
This sentence involves multiple triples, such as [Oxidative stress]-[related to]->[neuronal damage], [reactive oxygen species]-[damage]->[neurons], [Oxidative stress]-[damage]->[neurons], [Oxidative stress]-[accumulate]->[harmful reactive oxygen species]. 
Representing all these triples can be overwhelming. To tackle this issue, in the current implementation, we prompted GPT to annotate the most important triple from each sentence for validation with KG. However, these annotations are not always accurate and can lead to the oversight of critical knowledge.

\subsection{Expert Interview}

In addition to the three domain experts on our author team, we conducted interviews with two further experts (E1, E2) specializing in computational health informatics and medical research. 
\qianwen{E1 is a research scientist in medical informatics, holds a PhD degree, and has 8 years of experience in pharmacy and EHR data analysis. E2 is a PhD candidate in health informatics with 4 years of research experience. E2 has a strong clinical background with an MBBS degree, which is equivalent to an MD in the US.
}
These interviews aimed to obtain insights into the usability, effectiveness, and areas for improvement of \name. 
The two additional experts are not authors of this paper.
Each interview lasted approximately 60 minutes. 

\begin{itemize}[leftmargin=*, noitemsep]
    \item Introduction (10 minutes). We began by providing a background overview and demonstrating the various components and functionalities of \name. 
    \item Case Presentation (20 minutes). Following this introduction, we presented two specific use cases, Case 2 and Case 3 as detailed in \cref{subsec:usecase}. We also provided the responses from GPT-4 of the same questions for comparison. This segment was designed to familiarize the experts with \name's usage and work flow. During the demonstration, experts were encouraged to interrupt with questions and comments.
    \item Free Exploration (15 minutes). We allowed the experts 15 minutes of free exploration of the system, during which they were encouraged to think aloud and vocalize their thoughts.
    This interactive session was closely observed, with our team taking detailed notes on the experts' interactions with \name. 
    \item Discussion (15 minutes). The interview concluded with a semi-structured discussion, during which we collected feedback on five critical aspects: the accuracy of information provided, clarity of explanations, relevance of responses and recommendations, coverage of essential topics, and the system's overall ease of use.
\end{itemize}



\textbf{Usability.} 
Both experts commented that it was easy and intuitive to use the system without further assistance from the interviewers, stating all these features \textit{``basically have no learning curve involved''} (E1).
They were particularly impressed with \name's ability in elucidating the connections between entities and surfacing pertinent literature.
They also expressed that the progressive visualization and ext-step recommendations are useful features to guide exploration.

\textbf{Willing to Use.}
The experts emphasized their frequent reliance on evidence-based resources like UpToDate~\cite{UptoDate} for accessing drug and medical information, underscoring the importance of accuracy and credibility in their work. For instance, E1 highlighted a hesitancy to utilize GPT, \textit{``I rarely used GPT for those questions, due to concerns about its accuracy.''}
Both experts stated that tools like \name\, which integrate literature for supporting evidence, could dramatically shift their perspective on the utilization of AI-powered tools in their professional activities. 
The ability of \name\ to directly link to and leverage verified scientific articles not only bolsters the trustworthiness of the information provided but also aligns with the experts' existing practices of evidence-based verification~\cite{wang2023drava, yang2023harnessing}.



\textbf{Comparing \name\ with GPT-4.}
Both experts commented on the differences between \name\ and GPT.
First, compared with the well-structured response from \name, GPT responses are much longer and redundant, with several sentences repeating similar points.
For the same question \textit{``Which factors can trigger Alzheimer’s to get worse?''}, GPT responses include more than 10 factors while \name\ typical try to focus on the most important two or three factors.
Apart from the amount of the information, E1 also appreciated that the graph works as an effective scaffold to help track and organize the information, especially when dealing with multiple entities.
Second, E2 noted that GPT often employs an uncertain tone, for instance, characterizing a factor by saying it ``has been studied for its potential in slowing the progression of...'' Such phrasing might mitigate the risk of providing inaccurate responses, but tends to render the information ``less useful''(E2) in making decisions or gaining insights.
Third, both experts mentioned that highlighting relevant entities and relations in the text can help them effectively grasp the main points, offering a clear advantage over the undifferentiated text produced by GPT.

\textbf{Suggested Improvements.}
The experts offered insightful suggestions for enhancing \name.
E1 proposed allowing the integration of user-provided KGs into the system to help \name\ comprehensively identify reference articles and broaden the scope of supported evidences. 
E1 also suggested granting users the capability to evaluate the quality of the KGs integrated in \name.
These suggests are made considering the critical links between KG quality and the reliability of \name's outputs.
Meanwhile, E2 recommended providing users with the option to customize the verbosity and detail level of textual responses. 
This feature would accommodate varying user preferences and capacities for processing information, ensuring \name\ can cater to a diverse range of needs.

\section{Discussion}

\noindent
\textbf{Generality:} 
We developed and assessed \name\ within the domain of dietary supplements, a field where access to scientifically validated information is crucial due to widespread exaggerated claims and misinformation. 
Our evaluation highlights \name's efficacy in this particular context, and we are confident that our approach can be broadly applied to other areas. Firstly, \name\ is well-suited for domains where users typically use evidence-backed knowledge from scholarly literature for reasoning and decision making.
Secondly, it is applicable to information-seeking scenarios where data can be effectively organized as a graph, and the understanding can benefit from a step-by-step exploration.

\noindent
\textbf{Scalability:}
Even though the proposed method represents a significant improvement over traditional linear conversation methods, challenges may arise as the volume of content increases. 
Specifically, managing and navigating within a graph that contains large number of nodes and edges may become cumbersome for users. 
To tackle this problem, forthcoming improvements could aim at incorporating hierarchical structures and offering multiple abstraction levels. This would allow users to initially engage with a broad overview, delve into detailed information for areas of interest, and seamlessly transition between different abstraction layers.
Adaptive interface is another promising research direction for addressing this issue.
By learning the desired level of abstraction from interaction logs, the interface can dynamically adjust accordingly to suit the exploration needs of different users.


\noindent
\textbf{Inherent limitations related to KG:}
Inherent limitations associated with KGs must be acknowledged despite their effectiveness in representing structured knowledge. 
While KGs offer a powerful means of organizing and manipulating knowledge, they may not fully capture the complexity of human cognition, especially aspects that are less structured or context-dependent. In future research, we plan to investigate the integration of other formats of knowledge (\eg, images, tabluar data, maps) to further facilitate the information seeking with LLMs.

Meanwhile, it's crucial to recognize the information validation supported in \name\ is confined to the scope of the integrated KG. 
When a relation cannot be validated within the KG, it does not necessarily imply that the relation is untrue. 
As a result, users should be careful when interpreting the validation information derived from KGs. A promising future direction would be to dynamically integrate user knowledge into the system~\cite{wang2023drava, cheng2022polyphony}. 

\noindent
\qianwen{
\textbf{Further Enhancement of the Interface:}
While the effectiveness of the current  \name\ interface is demonstrated in our evaluation, its focus on intuitiveness and familiarity presents opportunities for further improvement of advanced feature in future studies. 
For instance, the current linear navigator could be expanded to include a tree-based option. This would allow users to more effectively track and compare different exploration paths related to a specific question. Additionally, the layout can be improved to a more semantic design that incorporates the embedding of graph nodes and supports focus + context interactions, which would facilitate the interpretation of larger and more complex knowledge relationships.
}

\noindent
\textbf{Limitations of the Evaluation: }
The evaluation of \name\ has limited user data, making it challenging to conduct quantitative assessments for both the whole system and the critical components such as entity matching and recommendation. 
To mitigate this limitation, our study emphasizes a close involvement of domain experts, gathering timely feedback during weekly meetings.
The use cases and the expert interview demonstrate the effectiveness and usability of the proposed system.
\qianwen{
At the same time, we recognize limitations in the current evaluation, such as restricted perspectives and limited generalizability. To address these issues, we plan to expand user participation in follow-up studies. Increasing the number of participants will enhance the robustness of our evaluation and provide a more comprehensive understanding of user behaviors through the analysis of chat texts and interaction logs.}

\section{Conclusion}
In conclusion, this paper presents \name, a visualization system that integrates LLMs with KGs to address the challenges of accuracy and structured exploration in health information seeking. 
\name\ tackles these issues of accuracy through the extraction and mapping of triples from LLM outputs to validated information in external KGs. 
It facilitates structured step-by-step exploration by providing recommendation based on KG neighborhood analysis, ensuring a comprehensive understanding without overlooking critical aspects.
Furthermore, to mitigate information overload during multi-step exploration, \name\ employs a focus+context design and introduces progressive graph visualization to track previous inquiries and connect them with current queries and next-step recommendations.

Our study demonstrates the effectiveness of \name\ in one critical application, dietary supplements, through use cases and expert interviews. 
We believe the proposed methods can be generalized to other similar application where structured exploration and validation through literature are essential.

\acknowledgments{    The authors thank the anonymous reviewers and the study participants for their valuable comments. 
	This work was partially supported by the National Institutes of Health’s National Center for Complementary and Integrative Health under grant number R01AT009457, National Institute on Aging under grant number R01AG078154 and National Cancer Institute under grant number R01CA287413. The content is solely the responsibility of the authors and does not represent the official views of the National Institutes of Health.
}

\bibliographystyle{abbrv-doi-hyperref}

\bibliography{template}








\end{document}